\documentclass[reprint,pra,twocolumn,longbibliography]{revtex4-1}
\usepackage[T1]{fontenc}
\usepackage{color}
\usepackage[colorlinks,urlcolor=cyan,citecolor=blue,linkcolor=magenta]{hyperref}
\usepackage{textcomp}
\usepackage{amsmath}
\usepackage{amssymb}
\usepackage{graphicx}
\usepackage{babel}

\begin{document}
\title{Dynamical generation of solitons in one-dimensional Fermi superfluids with and without spin-orbit coupling}

\author{Lingchii Kong$^{1,2}$, Genwang Fan$^{1}$, Shi-Guo Peng$^{4}$}

\author{Xiao-Long Chen$^{3}$}
\email{xiaolongchen@swin.edu.au}

\author{Huaisong Zhao$^{1}$}
\email{hszhao@qdu.edu.cn}

\author{Peng Zou$^{1}$ }
\email{phy.zoupeng@gmail.com}

\affiliation{$^{1}$College of Physics, Qingdao University, Qingdao 266071, China}
\affiliation{$^{2}$International Center for Quantum Materials, School of Physics,
Peking University, Beijing 100871, China}

\affiliation{$^{3}$Institute for Advanced Study, Tsinghua University, Beijing 100084, China}
\affiliation{$^{4}$State Key Laboratory of Magnetic Resonance and Atomic and Molecular
Physics, Innovation Academy for Precision Measurement Science and
Technology, Chinese Academy of Sciences, Wuhan 430071, China}
\date{June 9, 2021}

\begin{abstract}
We theoretically generalize a systematic language to describe the phase-imprinting technique to investigate the dynamical generation of solitons in a one-dimensional Raman-type spin-orbit-coupled Fermi superfluid. We check our method with the simulation of time-dependent Bogoliubov-de Gennes equations and find that our method not only can generate stable dark and even gray solitons in a conventional Fermi superfluid by controlling the transferred phase jump but also is feasible to create a stable dark soliton in both BCS and topological states of a spin-orbit-coupled Fermi superfluid. We also discuss the physical implication of our method.
\end{abstract}
\maketitle

\section{Introduction}

As an interesting nonlinear phenomenon, a soliton is a possible eigenstate of a many-body system. It originates from the competition between the dispersion and interaction of underlying systems and displays as a local and topological defect in the system~\cite{Drazin2002,Kevrekidis2008}. Solitons have also become the focus of research in ultracold atoms owing to their close relation to the dynamics of the system~\cite{Kevrekidis2008}. Their different form creates a large family, from the common gray and dark solitons in repulsive Bose-Einstein condensates (BECs) to the bright soliton in attractive BECs and gap soliton in optical lattices and also vector solitons, such as the bright-dark soliton in two-component BECs~\cite{Busch2001,Becker2008}. In fermionic superfluids, the soliton and its related dynamical behaviors have also attracted much research interest and have been widely investigated~\cite{Antezza2007,Robin2011,Spuntarelli2011,Liao2011,Robin2012njp,Efimkin2015}. After the realization of Raman-type spin-orbit-coupled (SOC) Fermi gases~\cite{Wang2012prl,Cheuk2012prl}, an exotic Majorana soliton joined this big family when the system comes into the topological state and displays a quite different dynamical behavior~\cite{Xu2014,Liu2015,Zou2016}.

Experimentally, solitons usually can be produced by a phase-imprinting technique or quench dynamics~\cite{Bongs2003,Burger1999} in BECs. In 2013, a "heavy soliton" in Fermi gases was observed and oscillated in a harmonic trap with a frequency surprisingly larger than theoretical prediction~\cite{Yefsah2013}. Later, people realized that the reason is the generated soliton decays into by-products, like vortex rings~\cite{Bulgac2014} and solitonic vortices~\cite{Donadello2014}, in the following experimental process. Since then, how to find a proper experimental strategy to generate a stable soliton in a Fermi superfluid has become a very interesting question and has attracted much research attention.

In 2014, Sacha and Delande first proposed a single-component operation of the phase-imprinting technique, which suggests that one should shine a phase laser beam on half of the gases and input a $\pi$ phase jump to only one spin component of the two-component Fermi superfluid in order to generate a stable soliton~\cite{Sacha2014}. This suggestion is surprising and quite interesting and opens the way to understand the dynamical generation of solitons. More important, it points out that the dynamical generation of a stable soliton should satisfy the parity symmetry of the soliton eigenstate, and the phase-imprinting technique is essentially a way to tune the parity of the wave function. Solitons can widely exist in many different kinds of systems, for example, the SOC Fermi superfluid~\cite{Xu2014,Liu2015} and dipolar gases~\cite{Pawlowski2015njp}. Naturally, it will also be interesting to investigate whether this single-component operation can work in other Fermi superfluids or not and to check the possibility to have a universal and systematic language of phase-imprinting techniques which can work in various Fermi superfluids.

In this paper, we will try to generalize the method of the phase-imprinting technique considering the parity symmetry of both ground and soliton states; we first study the dynamical generation of a gray soliton in a conventional Fermi superfluid following the relation between the soliton's velocity and its phase jump~\cite{Liao2011,Efimkin2015}, then analyze and understand the operation utilized in the soliton experiment~\cite{Yefsah2013} and explain the influence of the soliton's collision~\cite{Robin2012njp}, and finally study the dynamical generation of solitons in a SOC Fermi superfluid and discuss the physical meaning and experimental requirements of this operation. All simulations will be carried out with a time-dependent Bogoliubov-de Gennes (BdG) equation.

The rest of this paper is organized as follows. In the next section, we will introduce the model and Hamiltonian in both a conventional Fermi superfluid and a SOC Fermi superfluid and present the detailed process of the phase-imprinting method which meets the requirement of parity symmetry. In Sec.~\ref{sec3}, we investigate the time-dependent simulation of a stationary dark soliton and also a gray soliton in a conventional Fermi superfluid. Then we introduce the soliton's dynamical generation in all possible matter states of a Raman-type SOC Fermi superfluid in Sec.~\ref{sec4} and demonstrate the corresponding physical implication of the phase-imprinting technique in Sec.~\ref{sec5}. Finally, our conclusions are given in Sec.~\ref{sec6}.

\section{Model and Hamiltonian}
\subsection{One-dimensional conventional BCS Fermi superfluid}

Let us first review a one-dimensional (1D) conventional Fermi superfluid,
which can be considered a special case of the 1D Raman
SOC Fermi superfluid at the limit of zero Zeeman magnetic field. We
consider a uniform spin-balanced two-component Fermi superfluid with
an $s$-wave contact interaction at zero temperature, $T=0$.

In the frame of the mean-field theory,
all eigenstates of the system are described by the stationary BdG equations,
\begin{equation}
H_{\rm BdG}\Phi_{\eta}=E_{\eta}\Phi_{\eta}.
\label{eq:HBdG}
\end{equation}
Here the BdG Hamiltonian reads
\begin{equation}
H_{\rm BdG}=
\left[\begin{array}{cc}
\mathcal{H}_{s} & \Delta\\
\Delta^{*} & -\mathcal{H}_{s}
\end{array}\right],
\end{equation}
where $\mathcal{H}_{s}=-\partial_{x}^{2}/2m-\mu$ is a free-particle
Hamiltonian with atomic mass $m$ and chemical potential $\mu$. $\Phi_{\eta}=[u_{\eta},v_{\eta}]^T$ is
the quasiparticle wavefunction with the corresponding eigenenergy $E_{\eta}$.
Here and in the following, we always set $\hbar=1$ for simplicity. All
eigenstates of BdG equations should be self-consistently solved with the
order parameter equation
\begin{equation}
\Delta=-g_{{\rm 1D}}\sum_{\eta}u_{\eta}v_{\eta}^{*}f\left(E_{\eta}\right)
\end{equation}
and the density equation
\begin{equation}
n=2\sum_{\eta}\left[\left|u_{\eta}\right|^{2}f\left(E_{\eta}\right)+\left|v_{\eta}\right|^{2}f\left(-E_{\eta}\right)\right],
\end{equation}
where $f\left(x\right)=1/\left(e^{x/k_{B}T}+1\right)$ is the Fermi-Dirac
distribution function at temperature $T$.
The effective coupling
strength $g_{1D}<0$ can be described with a dimensionless interaction
strength parameter $\gamma=-mg_{{\rm 1D}}/n_{0}$, where $n_{0}$ is the
bulk density of the uniform system, and can often be used to define the non-interacting Fermi vector $k_F=\pi n_0/2$ and Fermi energy $\varepsilon_{F}=k_F^2/2m$. 
The typical value of the interaction strength is $\gamma \sim  3-5$ in a realistic experiment \cite{Liao2010,Liu2007pra,Liu2008pra}.
We have tried different $\gamma$ and the main conclusion is not changed. So in the following discussion we always take
$\gamma=\pi$.

Generally, the ground state of the system is a homogeneous state with an
even-parity symmetry of the order parameter $\Delta(x)$, namely,
$\Delta(-x)=\Delta(x)$. A soliton is an excited eigenstate with an odd-parity symmetry of the order parameter, namely, $\Delta(-x)=-\Delta(x)$.
The ground and soliton eigenstates have their own parity operators,
\begin{equation}
\begin{array}{cc}
P_{G}=\left[\begin{array}{cc}
1 & 0\\
0 & 1
\end{array}\right]P_{x}, & P_{S}=\left[\begin{array}{cc}
1 & 0\\
0 & -1
\end{array}\right]P_{x},\end{array}
\end{equation}
where $P_{x}$ is the usual parity operator, i.e., $P_{x}\Delta\left(x\right)\equiv\Delta\left(-x\right)$. $P_{G}$ and $P_{S}$ both commute with their corresponding $H_{{\rm BdG}}$, namely, $\left[P_{G/S},H_{{\rm BdG}}\right]=0$,
which means that the parities of these two states are conserved. One should notice that $H_{{\rm BdG}}$ of
these two eigenstates are not the same because of the parity difference
of their $\Delta\left(x\right)$. $P_G$ requires both quasiparticle wave functions $u_{\eta}$ and $v_{\eta}$
to be an even parity function, while $P_{S}$ requires the parity of $u_{\eta}$ to be different from that of $v_{\eta}$.

\subsection{One-dimensional Raman-type SOC Fermi superfluid }

For a 1D Raman-type SOC Fermi superfluid, the system can be described by the
model Hamiltonian $H=H_{0}+H_{{\rm int}}$, where
\begin{equation}
H_{0}=\int dx\left[\sum_{\sigma}\varPsi_{\sigma}^{\dagger}\mathcal{H}_{s}\varPsi_{\sigma}-h\left(\varPsi_{\uparrow}^{\dagger}e^{i2k_{R}x}\varPsi_{\downarrow}+H.c.\right)\right]
\end{equation}
is the single-particle Hamiltonian in the presence of a SOC effect, and
\begin{equation}
H_{{\rm int}}=g_{{\rm 1D}}\int dx\varPsi_{\uparrow}^{\dagger}\left(x\right)\varPsi_{\downarrow}^{\dagger}\left(x\right)\varPsi_{\downarrow}\left(x\right)\varPsi_{\uparrow}\left(x\right)
\end{equation}
is the $s$-wave contact interaction Hamiltonian
between two spin components. In the two-photon Raman SOC process, $h$
is the effective Zeeman magnetic field of the Raman beams, and $k_{R}$
is the recoil momentum carried by lasers.

It is useful to remove the
spatial dependence of the Raman coupling term by taking the following
local gauge transformation:
\begin{equation} \label{eq:u1}
\begin{array}{c}
\varPsi_{\uparrow}\left(x\right)=e^{+ik_{R}x}\widetilde{\psi}_{\uparrow}\left(x\right),\\
\varPsi_{\downarrow}\left(x\right)=e^{-ik_{R}x}\widetilde{\psi}_{\downarrow}\left(x\right);
\end{array} 
\end{equation}
here this unitary transformation keeps the physics of the spin index. The SOC effect in the Hamiltonian $H_{0}$ can be regarded as an equal-weight combination of Rashba and Dresselhaus spin-orbit couplings after the other unitary transformation
\begin{equation}
\begin{array}{c}
\widetilde{\psi}_{\uparrow}\left(x\right)=\frac{1}{\sqrt{2}}\left[\psi_{\uparrow}\left(x\right)-i\psi_{\downarrow}\left(x\right)\right],\\
\widetilde{\psi}_{\downarrow}\left(x\right)=\frac{1}{\sqrt{2}}\left[\psi_{\uparrow}\left(x\right)+i\psi_{\downarrow}\left(x\right)\right].
\end{array}\label{eq:u2}
\end{equation}
After the second transformation, we must emphasize that the spin indices on
the right side of Eq.~\eqref{eq:u2} 
do not denote original spin up or down. The corresponding single-particle
Hamiltonian reads
\begin{equation}
H_{0}=\int dx\left[\psi_{\uparrow}^{\dagger}\left(x\right),\psi_{\downarrow}^{\dagger}\left(x\right)\right]\mathcal{H}_{0}\left[\begin{array}{c}
\psi_{\uparrow}\left(x\right)\\
\psi_{\downarrow}\left(x\right)
\end{array}\right],
\end{equation}
with
\begin{equation}
\mathcal{H}_{0}=\mathcal{H}_{s}-h\sigma_{z}+\lambda\hat{k}_{x}\sigma_{y}.\label{eq:h0u2}
\end{equation}
Here a constant energy shift $E_{R}=k_{R}^{2}/2m$ is absorbed by
the chemical potential $\mu$, and $\hat{k}_{x}=-i\partial_{x}$ is
the momentum operator. $\lambda\equiv k_{R}/m$ is the SOC constant. $\sigma_{x,z}$ are Pauli's matrices. The form of the interaction Hamiltonian is invariant after the above
two unitary transformations, namely, $H_{{\rm int}}=g_{{\rm 1D}}\int dx\psi_{\uparrow}^{\dagger}\left(x\right)\psi_{\downarrow}^{\dagger}\left(x\right)\psi_{\downarrow}\left(x\right)\psi_{\uparrow}\left(x\right).$

In the frame of mean-field theory, we define an order parameter $\Delta\left(x\right)=-g_{{\rm 1D}}\left\langle \psi_{\downarrow}\left(x\right)\psi_{\uparrow}\left(x\right)\right\rangle $.
Then the interaction Hamiltonian is decoupled as
\begin{equation}
H_{{\rm int}}\simeq-\int dx\left[\Delta\psi_{\uparrow}^{\dagger}\psi_{\downarrow}^{\dagger}+\mathrm{H.c.}+\left|\Delta\right|^{2}/g_{{\rm 1D}}\right].
\end{equation}
By taking the Bogoliubov transformation to the field operator $\psi_{\sigma}\left(x\right)=\sum_{\eta}\left[u_{\sigma\eta}\left(x\right)c_{\eta}+v_{\sigma\eta}^{*}\left(x\right)c_{\eta}^{\dagger}\right],$ we
can transform the Hamiltonian into a non-interacting quasiparticle BdG
Hamiltonian with quasiparticle operators $c_{\eta}$ and $c_{\eta}^{\dagger}$,
which satisfies Eq.~\eqref{eq:HBdG}, but here
\begin{equation}
H_{{\rm BdG}}\equiv\left[\begin{array}{cccc}
\mathcal{H}_{s}-h & -i\lambda\hat{k}_{x} & 0 & -\Delta\\
i\lambda\hat{k}_{x} & \mathcal{H}_{s}+h & \Delta & 0\\
0 & \Delta^{*} & -\mathcal{H}_{s}+h & i\lambda\hat{k}_{x}\\
-\Delta^{*} & 0 & -i\lambda\hat{k}_{x} & -\mathcal{H}_{s}-h
\end{array}\right],
\end{equation}
and $\Phi_{\eta}=\left[u_{\uparrow\eta},u_{\downarrow\eta},v_{\uparrow\eta},v_{\downarrow\eta}\right]^{T}$
is the quasiparticle wave function with eigenenergy $E_{\eta}$. This BdG equation should also be self-consistently solved with the order parameter equation
\begin{equation}
\Delta=-\frac{g_{1D}}{2}\sum_{\eta}\left[u_{\uparrow\eta}v_{\downarrow\eta}^{*}f\left(E_{\eta}\right)+u_{\downarrow\eta}v_{\uparrow\eta}^{*}f\left(-E_{\eta}\right)\right]
\label{eq:socdel}
\end{equation}
and the density equation
\begin{equation}
n=\frac{1}{2}\sum_{\sigma\eta}\left[\left|u_{\sigma\eta}\right|^{2}f\left(E_{\eta}\right)+\left|v_{\sigma\eta}\right|^{2}f\left(-E_{\eta}\right)\right].
\label{eq:socden}
\end{equation}

\begin{figure}
\includegraphics[scale=0.4]{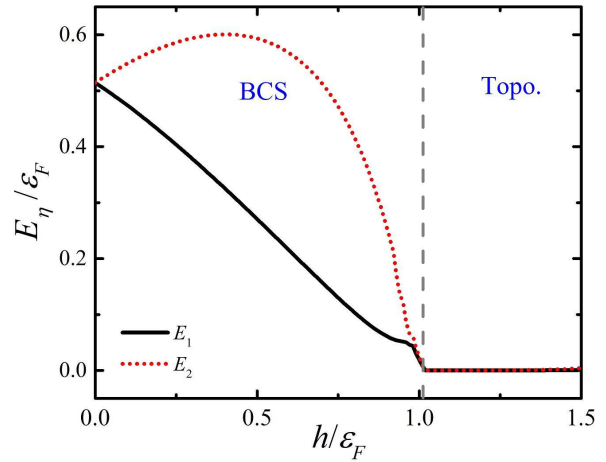}
\caption{\label{transition} The two lowest eigenenergies $E_1$ and $E_2$ of quasiparticles of a soliton state in a Raman-type SOC Fermi superfluid. The interaction strength $\gamma=\pi$, with recoil momentum $k_R=0.75k_F$. The critical Zeeman field from the BCS superfluid to the topological one is around $h_{c}\simeq\varepsilon_{F}$, across which $E_1$ and $E_2$ touch zero and the system
experiences a phase transition from a BCS superfluid to a topological
superfluid.}
\end{figure}

Similarly, the order parameter exhibits an even-parity symmetry with $\Delta\left(-x\right)=\Delta\left(x\right)$
in the ground state of this SOC system, while it has an odd-parity symmetry with $\Delta\left(-x\right)=-\Delta\left(x\right)$ in the soliton state. The parity operators of these two states respectively read
\begin{equation}
\begin{array}{cc}
P_{G}=\left[\begin{array}{cccc}
1 & 0 & 0 & 0\\
0 & -1 & 0 & 0\\
0 & 0 & -1 & 0\\
0 & 0 & 0 & 1
\end{array}\right]P_{x}, & P_{S}=\left[\begin{array}{cccc}
1 & 0 & 0 & 0\\
0 & -1 & 0 & 0\\
0 & 0 & 1 & 0\\
0 & 0 & 0 & -1
\end{array}\right]P_{x},\end{array}\label{eq:parity_soc}
\end{equation}
which both commute with their own $H_{{\rm BdG}}$ to conserve their parity property.
In the ground state, $P_{G}$ suggests that the quasiparticle wave functions $u_{\uparrow\eta}$ and $v_{\downarrow\eta}$
should be even-parity, while $u_{\downarrow\eta}$ and $v_{\uparrow\eta}$ are odd-parity, or the reverse since, mathematically, $-P_G$ is also a possible candidate ground-state parity operator. These requirements make $\Delta(x)$ to be an even function. In the soliton state, $P_S$ requires $u_{\uparrow\eta}$ and $v_{\uparrow\eta}$ to be even functions, while $u_{\downarrow\eta}$ and $v_{\downarrow\eta}$ are odd functions or vice versa. These requirements make
$\Delta(x)$ an odd function.

In our numerical simulation, we use the interaction strength $\gamma=\pi$, and recoil momentum $k_{R}=0.75k_{F}$. When increasing $h$ across a critical value $h_{c}\simeq\varepsilon_{F}$ (see Fig.~\ref{transition}), an interesting phase transition happens from a trivial BCS superfluid to a topologically nontrivial superfluid, and the critical transition point locates at the position where the values of the two lowest quasiparticle eigenenergies $E_1$ and $E_2$ in a soliton state just touch zero.

\subsection{Phase-imprinting strategy}

Experimentally, solitons can be produced by a phase-imprinting technique which can transform the system from the ground state $\Psi_{G}$ into the soliton state $\Psi_{S}$ (or the reverse). In this process, a far-detuning laser beam is shined on the system for a short time $dt$ to transfer a certain phase jump $\delta\phi$ to a chosen regime, i.e., the left part of the system in our discussion. Then a soliton will be generated on the edge between the left and right regimes \cite{Burger1999,Yefsah2013}. Mathematically, this process is defined by \begin{equation} \label{eq:imprintF}
F\Psi_{G}\equiv\Psi_{S}, 
\end{equation} 
where $F$ is a phase-imprinting operator. Although it is a local operator, it can globally change the parity property of the system. Playing the role of the unitary transformation matrix, $F$ can make the parity of $H_{{\rm BdG}}$ change as
\begin{equation}
F^{\dagger}H_{{\rm BdG}}\left[\Delta\right]F=H_{{\rm BdG}}\left[\Delta\rightarrow\Delta e^{-iI\left(x\right)}\right],\label{eq:ham_F}
\end{equation}
in which a local function $I\left(x\right)\equiv\Theta\left(-x\right)\delta\phi$
is defined to transfer a constant phase jump $\delta\phi$ to the
chosen part of the system.  $\Theta\left(x\right)$ is the Heaviside
step function. The matrix form of $F$ can be derived by the connection
of the parity operator of both the ground and soliton states,
\begin{equation}
P_{G}=P_{S}F.\label{eq:Fcompute}
\end{equation}
So once we know the parity operator $P_{G/S}$ in both the ground and soliton
states, we will immediately know the specific expression of the phase-imprinting
operator $F$.

Finally, a time-dependent version of the BdG equations
\begin{equation}
H_{{\rm BdG}}\Phi_{\eta}\left(t\right)=i\partial_{t}\Phi_{\eta}\left(t\right)
\end{equation}
 can be used to check the dynamical process of this phase-imprinting
operation. It should be self-consistently solved with Eqs.~\eqref{eq:socdel} and~\eqref{eq:socden}. Next, we will give the expression of $F$ in both cases
without and with the SOC effect.
\begin{figure}
\includegraphics[scale=0.4]{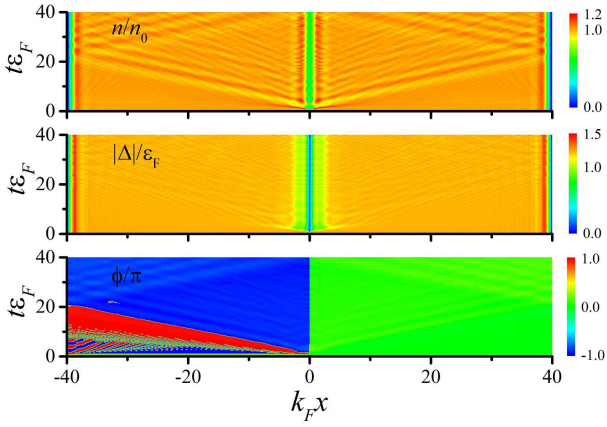}
\caption{\label{fignosoc} The density $n(x,t)$, amplitude $\left|\Delta\left(x,t\right)\right|$, and phase $\phi\left(x,t\right)$ of order parameter evolution of a 1D conventional BCS Fermi superfluid. A stable soliton is shown in the middle by inputting a $\delta\phi=\pi$ phase jump only to the left half of the quasiparticle wave function $v_{\eta}$. The system is transformed from a ground state into a soliton state at time $t\varepsilon_{F}=1$.}
\end{figure}

\section{Producing solitons in a conventional Fermi superfluid \label{sec3}}
\begin{figure}
\includegraphics[scale=0.4]{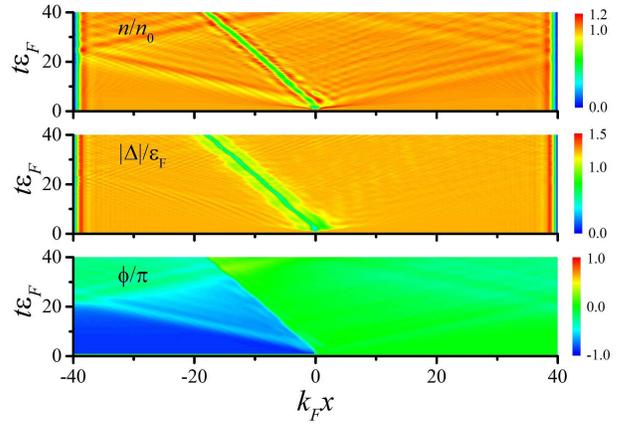}
\caption{\label{fignosoc-gray09} Same physical quantities as in Fig.~\ref{fignosoc} in the 1D conventional BCS Fermi superfluid. A left-moving gray soliton is generated at time $t\varepsilon_{F}=1$ in the middle by inputting a phase jump $\delta\phi=0.9\pi$ to only the left half of the quasiparticle wave function $v_{\eta}$.}
\end{figure}

We first discuss the case of a conventional Fermi superfluid without
a SOC effect. With Eq.~\eqref{eq:Fcompute}, it is easy to find that the
phase-imprinting operator $F$ reads
\begin{equation}
F=\left[\begin{array}{cc}
1 & 0\\
0 & -1
\end{array}\right]
\rightarrow
\left[\begin{array}{cc}
1 & 0\\
0 & e^{iI\left(x\right)}
\end{array}\right],
\end{equation}
where "$1$" in the above matrix means keeping the parity of the quasiparticle wave function $u_{\eta}$, while "$-1$" indicates that the phase-imprinting operation should be carried out to change the parity of $v_{\eta}$:  function $e^{iI\left(x\right)}$ shows the results that a phase jump $\delta\phi$ is transferred to the quasiparticle wave function $v_{\eta}$. Specifically $\delta\phi=\pi$ makes $e^{iI\left(x\right)}=-\Theta\left(-x\right)$, which is just the case in which we generate a dark soliton from the ground state. This $e^{iI\left(x\right)}$ will vary the parity of $v_{\eta}$ from an even parity to an odd one. The phase-imprinting operator $F$ will transform the quasiparticle wave functions following
\begin{equation}
\begin{array}{cc}
u_{\eta}\left(x,t+dt\right)= & u_{\eta}\left(x,t\right),\\
v_{\eta}\left(x,t+dt\right)= & e^{iI\left(x\right)}v_{\eta}\left(x,t\right).
\end{array}\label{eq:nosoc_vplan}
\end{equation}
Obviously, $-F$ is also a possible candidate phase-imprinting strategy, and it can produce the same results if we do a transformation to $u_{\eta}$ but not to $v_{\eta}$. The only difference between these two operations is that the sign of the phase jump should be different in order to generate the same soliton. This can be easily understood from the expression of the order parameter equation.

Next, a time-dependent simulation is used to check this dynamical operation.  Numerically, we take a box with length $k_FL=80$ to hold the system. The lowest 120 standing-wave bases are used to expand all quasiparticle wave functions, with an energy cutoff $E_c=25\varepsilon_{F}$. We have checked that a set of harsher calculation parameters will not qualitatively change our conclusions.

We first prepare a ground state and always carry out the phase-imprinting operation at time $t\varepsilon_{F}=1$. As shown in Fig.~\ref{fignosoc}, a stable dark soliton is clearly detected when transferring a phase jump $\delta\phi=\pi$. In the middle of the system where the soliton locates, the amplitude of the order parameter is zero the corresponding phase changes sign, and the phase jump is fixed at $\pi$ all the time. The generation of a soliton is accompanied by the transportation of sound wave ripples, which are induced by the density valley and Friedel oscillation. Besides a stationary soliton, it is known that the phase jump $\delta\phi$ can be used to control the speed of a soliton \cite{Liao2011,Efimkin2015}; a soliton can move by decreasing its phase jump $\delta\phi$. The faster the speed is, the smaller $\delta\phi$ is away from $\pi$. Following the same phase-imprinting strategy, we can input the phase jump $\delta\phi<\pi$ to the system by properly reducing the time duration of the phase-imprinting process. For example, when $\delta\phi=0.9\pi$, a gray soliton is generated successfully in Fig.~\ref{fignosoc-gray09}. The direction of the gray soliton is controlled by the sign of the phase jump $\delta\phi$, or the relative phase difference between the left and right parts of the system. This gray soliton moves towards left or right once we change the sign of the phase jump, namely, $\delta\phi=-0.9\pi$. Once the phase jump crosses a critical value and hinders the existence of a stable gray soliton, an unstable solitonlike product will be created and immediately decay into sound waves.
\begin{figure}
\includegraphics[scale=0.4]{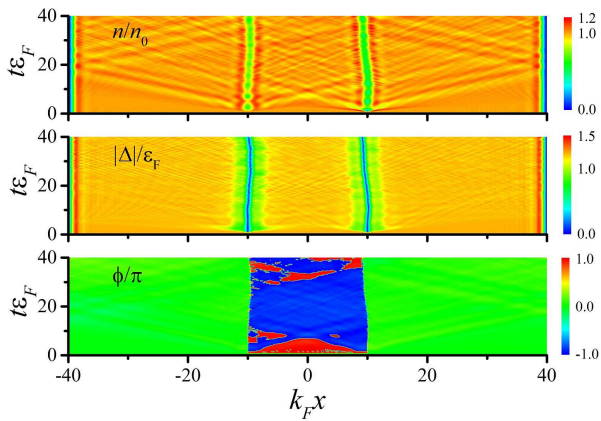}\caption{\label{fig:nosoc_2ds_10} Same physical quantities as in Fig.~\ref{fignosoc} in the 1D conventional BCS Fermi superfluid. Two dark solitons located at $k_{F}x=\pm10$ are generated by inputting a phase jump $\delta\phi=\pi$ to $u_{\eta}$ in the regime $k_{F}x<10$ and another phase jump $\delta\phi=\pi$ to $v_{\eta}$ in the regime $k_{F}x<-10$.}
\end{figure}

This interesting experimental strategy shown in Eq.~\eqref{eq:nosoc_vplan} was first suggested by Sacha and Delande \cite{Sacha2014} and was explained as a single-component phase-imprinting strategy because only the function $v_{\eta}$ is changed. Here $u_{\eta}$ and $v_{\eta}$ respectively denote the spin-up and spin-down quasiparticle wave functions \cite{Liu2007}. Physically, this single-component operation means that only spin-down atoms of Cooper pairs will be influenced by the phase laser beam, while the spin-up ones are not. Experimentally, an external potential, whose strength $I$ should be larger than the typical energy scale of the system to avoid the influence of other terms in the Hamiltonian, operates only on the spin-down component with a duration $dt$ shorter than any typical timescale of the system. By controlling the external potential strength and its duration, we can input a certain phase jump $Idt=\delta\phi$ to the system. To our knowledge, the experimental realization of the single-component operation in a Fermi superfluid is still not realized in the Fermi soliton experiment. Experimentally, this single-component strategy can potentially be realized by a tune-out wavelength technique \cite{Raisa2017,Jun2020}. Currently, the green laser beam used by the Massachusetts Institute of Technology (MIT) group is a dipole potential, which transfers phase variation to both spin components \cite{Yefsah2013}. This operation can be understood as two atoms in a Cooper pair both absorbing this phase jump $\delta\phi$ and varies the parity properties of both quasiparticle wave functions, $u_{\eta}$ and $v_{\eta}$. So here we call this operation a two-component operation.
\begin{figure}
\includegraphics[scale=0.4]{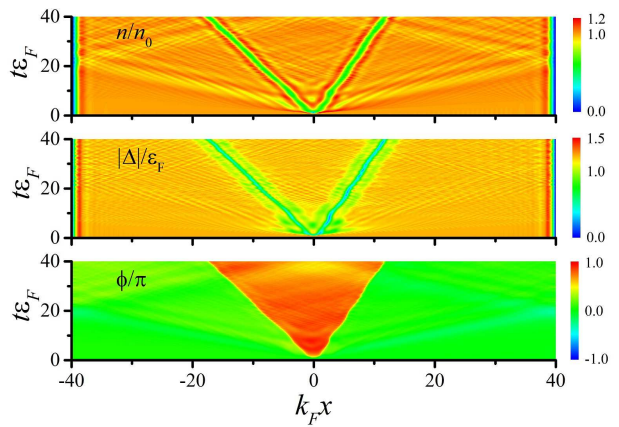}
\caption{\label{fignosoc-l09_r095} Same physical quantities as in Fig.~\ref{fignosoc} in the 1D conventional BCS Fermi superfluid. Two gray solitons with different speeds are generated by inputting a phase jump $\delta\phi= 0.95\pi$ to $u_{\eta}$ and phase jump $\delta\phi= 0.9\pi$ to $v_{\eta}$ at time $t\varepsilon_{F}=1$.}
\end{figure}

\begin{figure}
\includegraphics[scale=0.4]{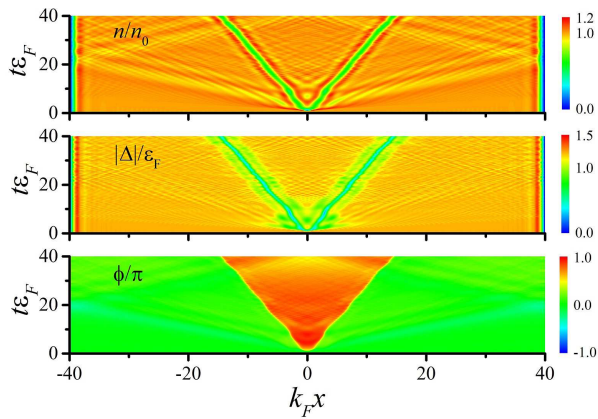}\caption{\label{fig:nosoc_2ds} Same physical quantities as in Fig.~\ref{fignosoc} in the 1D conventional BCS Fermi superfluid. A phase jump $\delta\phi=\pi$ is transferred to both $u_{\eta}$ and $v_{\eta}$ in the regime $k_Fx<0$; the interplay of two dark solitons makes them lose energy and decay into gray solitons.}
\end{figure}

Next, we will introduce the effect of the two-component operation. We first consider transferring a phase jump $\delta\phi=\pi$ to $v_{\eta}$ at position $k_{F}x<-10$ and another phase jump $\delta\phi=\pi$ to $u_{\eta}$ at $k_{F}x< 10$. As shown in Fig.~\ref{fig:nosoc_2ds_10}, two dark solitons are successfully detected at $k_{F}x=\pm10$, which are just the positions where phases of $u_{\eta}$ and $v_{\eta}$ jump, respectively. Then we consider the situation of transferring different phase jumps $\delta\phi$ to both $u_{\eta}$ and $v_{\eta}$ at the same position $k_Fx=0$. For example, we transfer phase jump $\delta\phi=0.95\pi$ to $u_{\eta}$ and phase jump $\delta\phi=0.9\pi$ to $v_{\eta}$. These two values of phase jumps are not too small and can support a stable gray soliton. The results are displayed in Fig.~\ref{fignosoc-l09_r095}, where two gray solitons with different velocities (different slopes) are detected. The left gray soliton has the same speed as the one in Fig.~\ref{fignosoc-gray09} since the same phase jump $\delta\phi=0.9\pi$ is transferred to $v_{\eta}$ (the same spin component). The velocity of the right gray soliton is smaller than that of the left one due to its larger amplitude of the phase jump. The results of both Figs.~\ref{fig:nosoc_2ds_10} and~\ref{fignosoc-l09_r095} indicate that operating on both spin components ($u_{\eta}$ and $v_{\eta}$) at the same time means the generation of two solitons. The different moving directions cause the two gray solitons to not have enough time to collide with each other. So the effect of this two-component operation can be understood as the combination effect when the single-component operations of each spin component are done separately.

What about the case in which we transfer phase jump $\delta\phi=\pi$ to both spin components at the same position, i.e., $k_Fx<0$? This operation is similar to what is carried out in the MIT experiment. As shown in Fig.~\ref{fig:nosoc_2ds}, instead of obtaining two overlapping stationary dark solitons, we observe two separate moving gray solitons. The strange dynamical behavior here is due to the collision of two solitons \cite{Robin2012njp}. The whole dynamical process can be understood as follows: initially transferring phase jump $\delta\phi=\pi$ to both spin components produces two overlapping stationary solitons with zero velocity, which allows them to have enough time to collide with each other. The inelastic collision induces solitons to lose energy and obtain a velocity to avoid sharing the same spatial location and finally makes solitons evolve into gray solitons. We have checked that the final velocity of solitons here is influenced by the interaction strength $\gamma$. A bigger interaction strength $\gamma$ will generate faster gray solitons.

To end this section, let us consider the special case of transferring a phase jump $\delta\phi=-\pi/2$ to $u_{\eta}$ and another phase jump $\delta\phi=\pi/2$ to $v_{\eta}$ at $k_Fx<0$. Initially, this operation can induce an odd-parity symmetry of $\Delta(x)$, which looks like the requirements of a stable dark soliton. However, based on our above discussion, it is not difficult to see that this is impossible. Usually, $|\delta\phi|=\pi/2$ means a very small phase jump, in which solitons have already decayed into sound waves. Also, the same value but different sign of the phase jump means the generation of two overlapping solitons whose inelastic collision will make them lose energy and speed up their decay process. The discussion above indicates that the parity requirements of $u_{\eta}$ and $v_{\eta}$ are a sufficient condition to obtain a ground state or soliton state, while the parity requirements of $\Delta(x)$ are just a necessary condition.

\section{Producing solitons in a SOC Fermi superfluid \label{sec4}}

Now we discuss the dynamical generation of a soliton in a Raman-type SOC Fermi superfluid at different Zeeman magnetic fields $h$. With the parity operators of both the ground and soliton states in Eq.~\eqref{eq:parity_soc} and following a derivation similar to Eq.~\eqref{eq:Fcompute}, it is also easy to find that the phase-imprinting operator $F$ of the SOC Fermi superfluid should take the expression
\begin{equation}
F=\left[\begin{array}{cccc}
1 & 0 & 0 & 0\\
0 & 1 & 0 & 0\\
0 & 0 & -1 & 0\\
0 & 0 & 0 & -1
\end{array}\right]
\rightarrow
\left[\begin{array}{cccc}
1 & 0 & 0 & 0\\
0 & 1 & 0 & 0\\
0 & 0 & e^{iI\left(x\right)} & 0\\
0 & 0 & 0 & e^{iI\left(x\right)}
\end{array}\right].\label{eq:F}
\end{equation}
\begin{figure}[t]
\includegraphics[scale=0.4]{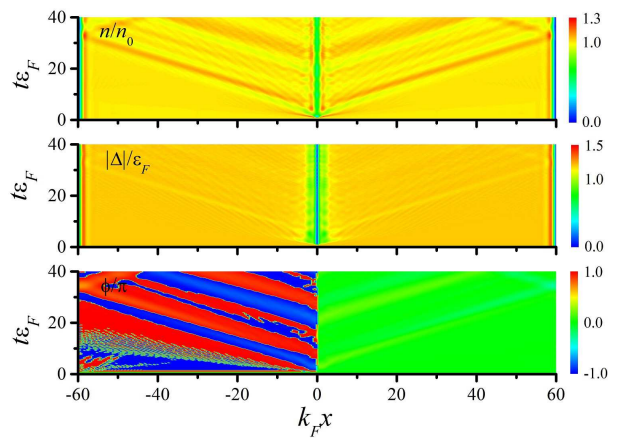}
\caption{\label{fig2_h03} Same physical quantities as in Fig.~\ref{fignosoc} in the Raman-type SOC Fermi superfluid in the BCS state ($h=0.3\varepsilon_{F}$). With the phase-imprinting operation described by Eq.~\eqref{eq:soc_plan1}, the system is transformed from a ground state into a soliton state at time $t\varepsilon_{F}=1$.}
\end{figure}
This expression indicates that a soliton can be produced by transferring a phase jump $\delta\phi$ to quasiparticle wave functions $v_{\uparrow\eta}$ and $v_{\downarrow\eta}$ of the ground state while retaining the parity of the other quasiparticle wave functions, $u_{\uparrow\eta}$ and $u_{\downarrow\eta}$. Specifically, phase jump $\delta\phi=\pi$ makes $e^{iI\left(x\right)}=-\Theta\left(-x\right)$; then we can generate a dark soliton from the ground state. The transformation of quasiparticle wave functions $F\Phi_{\eta}$ reads
\begin{equation}
\begin{array}{cc}
u_{\uparrow\eta}\left(x,t+dt\right)= & u_{\uparrow\eta}\left(x,t\right),\\
u_{\downarrow\eta}\left(x,t+dt\right)= & u_{\downarrow\eta}\left(x,t\right),\\
v_{\uparrow\eta}\left(x,t+dt\right)= & e^{iI\left(x\right)}v_{\uparrow\eta}\left(x,t\right),\\
v_{\downarrow\eta}\left(x,t+dt\right)= & e^{iI\left(x\right)}v_{\downarrow\eta}\left(x,t\right).
\end{array}\label{eq:soc_plan1}
\end{equation}
Next, following this phase-imprinting operation, we will use time-dependent BdG equations to investigate the dynamical generation of a soliton in both a BCS superfluid and a topological superfluid. Numerically, we take a box with length $k_FL=120$ to hold the system. The lowest 150 standing-wave bases are used to expand all quasiparticle wave functions, with an energy cutoff $E_c=25\varepsilon_{F}$. We have checked that a set of harsher calculation parameters will not qualitatively change our conclusions. We use the same physical parameters as in Fig.~\ref{transition}. Here we just discuss the generation of a dark soliton with phase jump $\delta\phi=\pi$. Before introducing the physical meaning of Eq.~\eqref{eq:soc_plan1}, let us check its validity.
\begin{figure}
\includegraphics[scale=0.4]{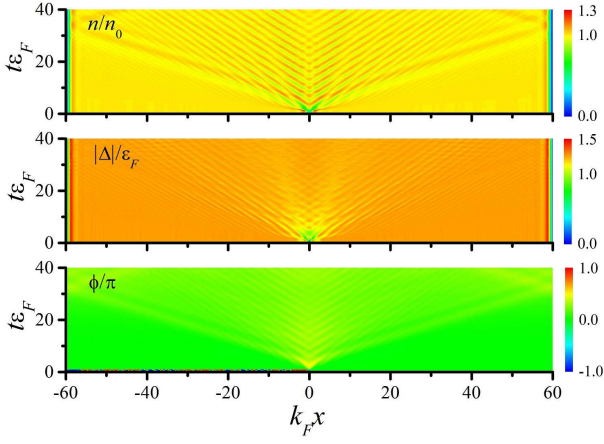}
\caption{\label{fig2_h03_inv} Same physical quantities as in Fig.~\ref{fignosoc} in the Raman-type SOC Fermi superfluid in the BCS state ($h=0.3\varepsilon_{F}$). With the phase-imprinting operation described by Eq.~\eqref{eq:soc_plan1}, the system is transformed from a soliton state into a ground state at time $t\varepsilon_{F}=1$.}
\end{figure}

We initially prepare a ground state, and then shine a phase-imprinting laser beam on the system at the time $t\varepsilon_{F}=1$. In the BCS superfluid with a Zeeman magnetic field $h=0.3\varepsilon_{F}$. As shown in Fig.~\ref{fig2_h03}, a dark soliton is successfully created in the middle, where a density valley and a zero mode of the order parameter are both clearly detected, with which we find the location of the soliton. When time goes on, the soliton is stable, and the amplitude of its order parameter is kept to zero, while the phase jump is fixed to $\pi$ all the time. In this process, some accompanying density stripes are also generated because the generation of a density valley will push some particles away from the soliton. Also, our operation does not simulate exactly the Friedel oscillation. This operation brings some extra energy to the system, which has to be released in the form of sound waves. Inversely, if we initially prepare a soliton state and repeat the same operation on the system, as shown in Fig.~\ref{fig2_h03_inv}, we can also destroy a soliton, in the process of which some accompanying density sound waves are also produced.

\begin{figure}
\includegraphics[scale=0.4]{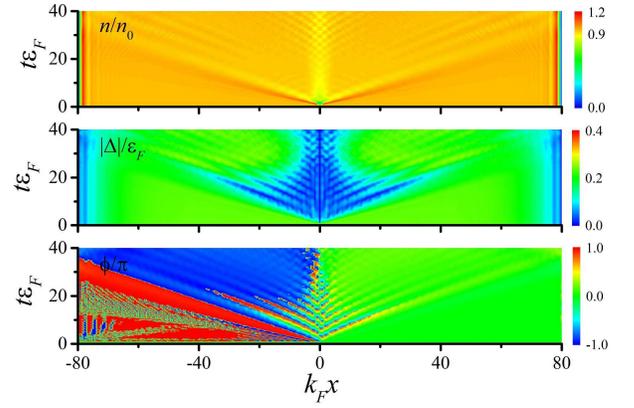}
\caption{\label{fig2_h11}Same physical quantities as in Fig.~\ref{fignosoc} in the Raman-type SOC Fermi superfluid in the topological state ($h=1.1\varepsilon_{F}$). With the phase-imprinting operation described by Eq.~\eqref{eq:soc_plan1}, the system is transformed from a ground state into a Majorana soliton state at time $t\varepsilon_{F}=1$.}
\end{figure}

\begin{figure}
\includegraphics[scale=0.4]{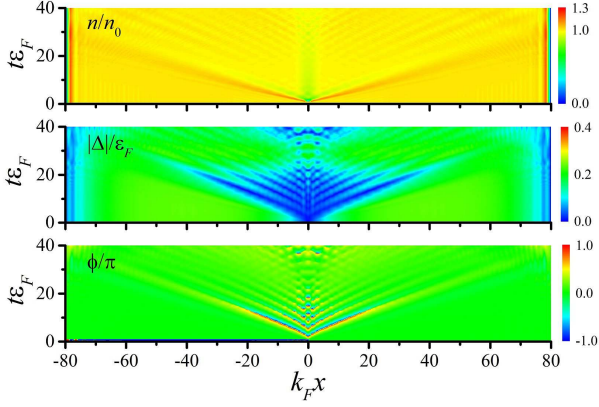}
\caption{\label{fig2_h11_inv}Same physical quantities as in Fig.~\ref{fignosoc} in the Raman-type SOC Fermi superfluid in the topological state ($h=1.1\varepsilon_{F}$). With the phase-imprinting operation described by Eq.~\eqref{eq:soc_plan1}, the system is transformed from a Majorana soliton state into a ground state at time $t\varepsilon_{F}=1$.}
\end{figure}

At a Zeeman magnetic field $h=1.1\varepsilon_{F}$, the system becomes a topological superfluid. Different from the soliton in a BCS superfluid, there is no density valley in a Majorana soliton \cite{Xu2014,Liu2015}. In fact, a Majorana soliton is very special and can move while fixing the value of the phase jump to $\pi$ \cite{Zou2016}. The location of a Majorana soliton can be detected by its zero amplitude of the order parameter or the transition position of the phase of the order parameter. In Fig.~\ref{fig2_h11}, our phase-imprinting operation successfully creates a Majorana soliton in the middle, across which there is a constant phase jump $\delta\phi=\pi$. Furthermore, some other accompanying Majorana solitons are generated by the Friedel oscillation, transported towards both the left and right, because the motion of a Majorana soliton does not need to vary the phase jump. However, a Majorana soliton has a critical speed over which itn will be unstable \cite{Zou2016}. This instability is the reason why, as time goes on, these Majorana solitons will gradually decay into sound wave ripples. Finally, only the middle Majorana soliton produced by our phase-imprinting strategy is left. Inversely, when we start from a Majorana soliton state, and repeat the same phase-imprinting operation on the system, as displayed in Fig.~\ref{fig2_h11_inv}, the middle Majorana soliton can also be destroyed, while the accompanying Majorana soliton is still generated and decays into sound waves.

In summary, the phase-imprinting strategy in Eq.~\eqref{eq:soc_plan1} can produce and destroy a soliton successfully in both BCS and topological superfluids. In fact, reducing the phase jump can also help to control the speed of the soliton, allowing us to generate a gray soliton in BCS state. In the topological state, however, how to control the speed of a Majorana soliton is an interesting, but unsolved, question which needs more research attention.

\section{Physical implication of the operation "F" \label{sec5}}

Now we discuss the physical meaning of the phase-imprinting operation in Eq.~\eqref{eq:soc_plan1} and theoretically introduce a two-step operation to realize this operation. Since the spin indices of the quasiparticle wave function in Eq.~\eqref{eq:soc_plan1} have lost their original physics, it is better to explain the principle of operation with the Hamiltonian and wave function after only the first local gauge transformation [Eq.~\eqref{eq:u1}], where the spin indices keep their original physics. We obtain the BdG Hamiltonian
\begin{equation}
\widetilde{H}_{{\rm BdG}}\equiv\left[\begin{array}{cccc}
\mathcal{H}_{s}+\lambda\hat{k}_{x} & -h & 0 & -\widetilde{\Delta}\\
-h & \mathcal{H}_{s}-\lambda\hat{k}_{x} & \widetilde{\Delta} & 0\\
0 & \widetilde{\Delta}^{*} & -\mathcal{H}_{s}+\lambda\hat{k}_{x} & h\\
-\widetilde{\Delta}^{*} & 0 & h & -\mathcal{H}_{s}-\lambda\hat{k}_{x}
\end{array}\right]
\end{equation}
in the mean-field frame after a similar Bogoliubov transformation $\widetilde{\psi}_{\sigma}\left(x\right)=\sum_{\eta}\left[\widetilde{u}_{\sigma\eta}\left(x\right)c_{\eta}+\widetilde{v}_{\sigma\eta}^{*}\left(x\right)c_{\eta}^{\dagger}\right]$ to $\widetilde{\psi}_{\uparrow}$ and $\widetilde{\psi}_{\downarrow}$. With the second unitary transformation in Eq.~\eqref{eq:u2}, it is easy to get the relation of quasiparticle wave functions after only one and two unitary transformations,
\begin{equation}
\left[\begin{array}{c}
\widetilde{u}_{\uparrow\eta}\\
\widetilde{u}_{\downarrow\eta}\\
\widetilde{v}_{\uparrow\eta}\\
\widetilde{v}_{\downarrow\eta}
\end{array}\right]=\frac{1}{\sqrt{2}}\left[\begin{array}{cccc}
1 & -i & 0 & 0\\
1 & i & 0 & 0\\
0 & 0 & 1 & i\\
0 & 0 & 1 & -i
\end{array}\right]\left[\begin{array}{c}
u_{\uparrow\eta}\\
u_{\downarrow\eta}\\
v_{\uparrow\eta}\\
v_{\downarrow\eta}
\end{array}\right],\label{eq:relation}
\end{equation}
with which we find the phase-imprinting strategy in Eq.~\eqref{eq:soc_plan1}
is mathematically equivalent to
\begin{equation}
\begin{array}{cc}
\widetilde{u}_{\uparrow\eta}\left(x,t+dt\right)= & \widetilde{u}_{\uparrow\eta}\left(x,t\right),\\
\widetilde{u}_{\downarrow\eta}\left(x,t+dt\right)= & \widetilde{u}_{\downarrow\eta}\left(x,t\right),\\
\widetilde{v}_{\uparrow\eta}\left(x,t+dt\right)= & e^{iI\left(x\right)}\widetilde{v}_{\uparrow\eta}\left(x,t\right),\\
\widetilde{v}_{\downarrow\eta}\left(x,t+dt\right)= & e^{iI\left(x\right)}\widetilde{v}_{\downarrow\eta}\left(x,t\right).
\end{array}\label{eq:soc_plan1-u1}
\end{equation}
Equations~\eqref{eq:relation} and~\eqref{eq:soc_plan1-u1} clearly indicate that the phase-imprinting operation should be carried out on both quasiparticle wave functions $\widetilde{v}_{\uparrow\eta}$ and $\widetilde{v}_{\downarrow\eta}$, which means transferring the phase jump to both spin components. This two-component operation is different from the single-component operation in a conventional Fermi superfluid. And this operation makes $\widetilde{H}_{{\rm BdG}}$ exactly meet the requirement in Eq.~\eqref{eq:ham_F}.
\begin{figure}
\includegraphics[scale=0.4]{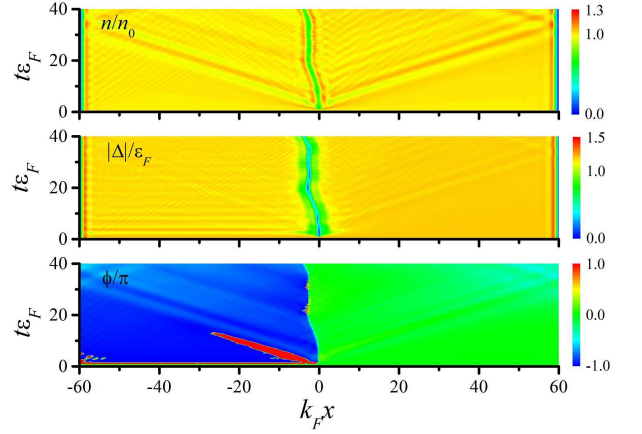}
\caption{\label{figbcs_1c}Same physical quantities as in Fig.~\ref{fignosoc} in the Raman-type SOC Fermi superfluid. This dynamical simulation with only a single-component phase-imprinting strategy $F_{1}$ fails to produce a stable soliton  in the BCS state ($h=0.3\varepsilon_{F}$).}
\end{figure}

Physically, the realization of the phase-imprinting operation $F$ mentioned above requires a two-step operation to tune the Hamiltonian. Mathematically, we denote this process by the equation $F=F_{2}F_{1}$. The first operation, $F_{1}$, is the same single-component operation which has been discussed in the conventional Fermi superfluid, and the second operator, $F_{2}$, is related to proper control of the Zeeman magnetic field $h$. In the Raman-type SOC Fermi superfluid, the expression of the single-component operation $F_{1}$ is
\begin{equation}
F_{1}=\left[\begin{array}{cccc}
e^{-iI\left(x\right)} & 0 & 0 & 0\\
0 & 1 & 0 & 0\\
0 & 0 & e^{iI\left(x\right)} & 0\\
0 & 0 & 0 & 1
\end{array}\right],
\end{equation}
where a phase jump $-\delta\phi$ is transferred to $u_{\uparrow\eta}$, while $\delta\phi$ transfers to $v_{\uparrow\eta}$, which can be realized by a common potential of spin-up atoms. However, this operation itself transforms $\widetilde{H}_{{\rm BdG}}$ as
\begin{equation}
\begin{array}{c}
F_{1}^{\dagger}\widetilde{H}_{{\rm BdG}}F_{1}\\
\equiv\left[\begin{array}{cccc}
\mathcal{H}_{s}+\lambda\hat{k}_{x} & -he^{iI\left(x\right)} & 0 & -\widetilde{\Delta}e^{iI\left(x\right)}\\
-he^{-iI\left(x\right)} & \mathcal{H}_{s}-\lambda\hat{k}_{x} & \widetilde{\Delta}e^{iI\left(x\right)} & 0\\
0 & \widetilde{\Delta}^{*}e^{-iI\left(x\right)} & -\mathcal{H}_{s}+\lambda\hat{k}_{x} & he^{-iI\left(x\right)}\\
-\widetilde{\Delta}^{*}e^{-iI\left(x\right)} & 0 & he^{iI\left(x\right)} & -\mathcal{H}_{s}-\lambda\hat{k}_{x}
\end{array}\right],
\end{array}
\end{equation}
\begin{figure}
\includegraphics[scale=0.4]{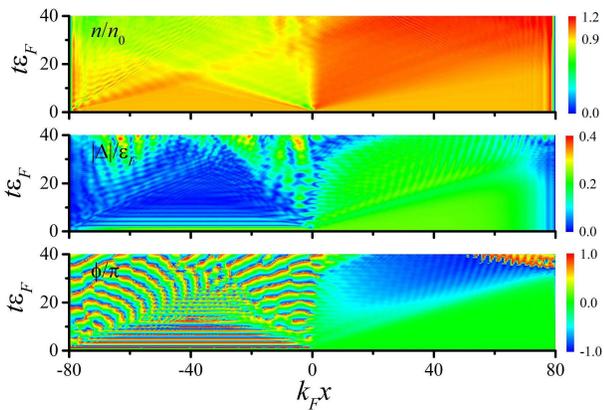}
\caption{\label{figtp_1c}Same physical quantities as in Fig.~\ref{fignosoc} in the Raman-type SOC Fermi superfluid. This dynamical simulation with only a single-component phase-imprinting strategy $F_{1}$ fails to produce a stable soliton  in the topological state ($h=1.1\varepsilon_{F}$).}
\end{figure}
which fails to meet the requirement shown in Eq.~\eqref{eq:ham_F} with some phase variations $e^{\pm iI(x)}$ brought alongside the order parameter $\tilde{\Delta}$. As shown by the time-dependent simulation in Figs.~\ref{figbcs_1c} and~\ref{figtp_1c}, this single-component operation $F_{1}$ fails to generate a stable dark soliton in both the BCS superfluid and topological superfluid when $\delta\phi=\pi$. The signal in the BCS superfluid bends like a snake and presents almost no obvious signal of a Majorana soliton in the topological superfluid. In order to cancel the influence of the Zeeman magnetic field $h$, the second operation, $F_{2}$,
\begin{equation}
F_{2}=\left[\begin{array}{cccc}
e^{iI\left(x\right)} & 0 & 0 & 0\\
0 & 1 & 0 & 0\\
0 & 0 & 1 & 0\\
0 & 0 & 0 & e^{iI\left(x\right)}
\end{array}\right],
\end{equation}
is carried out and makes $\widetilde{H}_{{\rm BdG}}$ satisfy
\begin{equation}
\begin{array}{c}
F_{2}^{\dagger}\widetilde{H}_{{\rm BdG}}F_{2}\\
\equiv\left[\begin{array}{cccc}
\mathcal{H}_{s}+\lambda\hat{k}_{x} & -he^{-iI\left(x\right)} & 0 & -\widetilde{\Delta}\\
-he^{iI\left(x\right)} & \mathcal{H}_{s}-\lambda\hat{k}_{x} & \widetilde{\Delta} & 0\\
0 & \widetilde{\Delta}^{*} & -\mathcal{H}_{s}+\lambda\hat{k}_{x} & he^{iI\left(x\right)}\\
-\widetilde{\Delta}^{*} & 0 & he^{-iI\left(x\right)} & -\mathcal{H}_{s}-\lambda\hat{k}_{x}
\end{array}\right],
\end{array}
\end{equation}
which can help to cancel the phase variation alongside the order parameter $\tilde{\Delta}$ and its conjugate. Operation $F_{2}$ can be realized by utilizing the SOC term in the Hamiltonian to transfer the phase jump to the effective Zeeman magnetic field $h$: during the process of annihilating a spin-down atom and generating a spin-up one, a phase jump $\delta\phi$ is transferred to $h$; inversely, annihilating a spin-up atom and generating a spin-down one, a phase jump $\delta\phi$ is transferred to $h$. In this process, the Hermiticity of the Hamiltonian is satisfied. As a special case, one just needs to change the sign of the Zeeman magnetic field $h$ in the left half during the phase-imprinting process to produce a stable dark soliton. The effect of these two operations together is just the phase-imprinting operation we need to produce a stable soliton in the Raman-type SOC Fermi superfluid.

\section{Conclusions \label{sec6}}

In summary, we theoretically generalized a systematic language to describe the phase-imprinting operation based on parity operators of both ground and soliton states which can be used to create a stable soliton in a one-dimensional Raman-type spin-orbit-coupled Fermi superfluid. The physical implication of this operation in spin-orbit-coupled a Fermi superfluid was also discussed. Based on time-dependent simulation of Bogoliubov-de Gennes equations, we found our suggestions can produce not only a stable dark or gray soliton by controlling the phase jump to the system in a conventional Fermi superfluid but also a stable dark soliton in both a BCS superfluid and a topological superfluid. Although all simulations were carried out with a mean-field theory whose prediction is often quantitatively inaccurate in low dimensions, we expect Bogoliubov-de Gennes equations could provide a qualitatively correct result. All discussions were limited to one dimension with zero temperature; the same idea can be generalized to other higher dimensions or different systems.

\begin{acknowledgments}
We are grateful for fruitful discussions with Hui Hu. This research was supported by the National Natural Science Foundation of China, Grants No. 11804177 (P.Z.), No. 11547034 (H.Z.), No. 11974384 (S.-G.P.); NKRDP under Grant No. 2016YFA0301503 (S.-G.P.); China Postdoctoral Science Foundation, Grant No. 2020M680495 (X.-L.C.); and the Shandong Provincial Natural Science Foundation, China, Grant No. ZR2018BA032 (P.Z.).
\end{acknowledgments}

\bibliography{solitonSOC}

\providecommand{\noopsort}[1]{}\providecommand{\singleletter}[1]{#1}%
\begin{thebibliography}{28}%
\makeatletter
\providecommand \@ifxundefined [1]{%
 \@ifx{#1\undefined}
}%
\providecommand \@ifnum [1]{%
 \ifnum #1\expandafter \@firstoftwo
 \else \expandafter \@secondoftwo
 \fi
}%
\providecommand \@ifx [1]{%
 \ifx #1\expandafter \@firstoftwo
 \else \expandafter \@secondoftwo
 \fi
}%
\providecommand \natexlab [1]{#1}%
\providecommand \enquote  [1]{``#1''}%
\providecommand \bibnamefont  [1]{#1}%
\providecommand \bibfnamefont [1]{#1}%
\providecommand \citenamefont [1]{#1}%
\providecommand \href@noop [0]{\@secondoftwo}%
\providecommand \href [0]{\begingroup \@sanitize@url \@href}%
\providecommand \@href[1]{\@@startlink{#1}\@@href}%
\providecommand \@@href[1]{\endgroup#1\@@endlink}%
\providecommand \@sanitize@url [0]{\catcode `\\12\catcode `\$12\catcode
  `\&12\catcode `\#12\catcode `\^12\catcode `\_12\catcode `\%12\relax}%
\providecommand \@@startlink[1]{}%
\providecommand \@@endlink[0]{}%
\providecommand \url  [0]{\begingroup\@sanitize@url \@url }%
\providecommand \@url [1]{\endgroup\@href {#1}{\urlprefix }}%
\providecommand \urlprefix  [0]{URL }%
\providecommand \Eprint [0]{\href }%
\providecommand \doibase [0]{http://dx.doi.org/}%
\providecommand \selectlanguage [0]{\@gobble}%
\providecommand \bibinfo  [0]{\@secondoftwo}%
\providecommand \bibfield  [0]{\@secondoftwo}%
\providecommand \translation [1]{[#1]}%
\providecommand \BibitemOpen [0]{}%
\providecommand \bibitemStop [0]{}%
\providecommand \bibitemNoStop [0]{.\EOS\space}%
\providecommand \EOS [0]{\spacefactor3000\relax}%
\providecommand \BibitemShut  [1]{\csname bibitem#1\endcsname}%
\let\auto@bib@innerbib\@empty
\bibitem [{\citenamefont {Drazin}\ and\ \citenamefont
  {Johnson}(2002)}]{Drazin2002}%
  \BibitemOpen
  \bibfield  {author} {\bibinfo {author} {\bibfnamefont {Philip~G}\
  \bibnamefont {Drazin}}\ and\ \bibinfo {author} {\bibfnamefont {Robin~S}\
  \bibnamefont {Johnson}},\ }\href@noop {} {\emph {\bibinfo {title} {Solitons:
  an introduction}}}\ (\bibinfo  {publisher} {Cambridge university press},\
  \bibinfo {year} {2002})\BibitemShut {NoStop}%
\bibitem [{\citenamefont {Carr}\ and\ \citenamefont
  {Brand}(2008)}]{Kevrekidis2008}%
  \BibitemOpen
  \bibfield  {author} {\bibinfo {author} {\bibfnamefont {Lincoln~D}\
  \bibnamefont {Carr}}\ and\ \bibinfo {author} {\bibfnamefont {Joachim}\
  \bibnamefont {Brand}},\ }\href@noop {} {\enquote {\bibinfo {title} {Emergent
  nonlinear phenomena in bose-einstein condensates},}\ } (\bibinfo {year}
  {2008})\BibitemShut {NoStop}%
\bibitem [{\citenamefont {Busch}\ and\ \citenamefont
  {Anglin}(2001)}]{Busch2001}%
  \BibitemOpen
  \bibfield  {author} {\bibinfo {author} {\bibfnamefont {Th.}\ \bibnamefont
  {Busch}}\ and\ \bibinfo {author} {\bibfnamefont {J.~R.}\ \bibnamefont
  {Anglin}},\ }\bibfield  {title} {\enquote {\bibinfo {title} {Dark-bright
  solitons in inhomogeneous bose-einstein condensates},}\ }\href {\doibase
  10.1103/PhysRevLett.87.010401} {\bibfield  {journal} {\bibinfo  {journal}
  {Phys. Rev. Lett.}\ }\textbf {\bibinfo {volume} {87}},\ \bibinfo {pages}
  {010401} (\bibinfo {year} {2001})}\BibitemShut {NoStop}%
\bibitem [{\citenamefont {Becker}\ \emph {et~al.}(2008)\citenamefont {Becker},
  \citenamefont {Stellmer}, \citenamefont {Soltan-Panahi}, \citenamefont
  {D{\"o}rscher}, \citenamefont {Baumert}, \citenamefont {Richter},
  \citenamefont {Kronj{\"a}ger}, \citenamefont {Bongs},\ and\ \citenamefont
  {Sengstock}}]{Becker2008}%
  \BibitemOpen
  \bibfield  {author} {\bibinfo {author} {\bibfnamefont {Christoph}\
  \bibnamefont {Becker}}, \bibinfo {author} {\bibfnamefont {Simon}\
  \bibnamefont {Stellmer}}, \bibinfo {author} {\bibfnamefont {Parvis}\
  \bibnamefont {Soltan-Panahi}}, \bibinfo {author} {\bibfnamefont {S{\"o}ren}\
  \bibnamefont {D{\"o}rscher}}, \bibinfo {author} {\bibfnamefont {Mathis}\
  \bibnamefont {Baumert}}, \bibinfo {author} {\bibfnamefont {Eva-Maria}\
  \bibnamefont {Richter}}, \bibinfo {author} {\bibfnamefont {Jochen}\
  \bibnamefont {Kronj{\"a}ger}}, \bibinfo {author} {\bibfnamefont {Kai}\
  \bibnamefont {Bongs}}, \ and\ \bibinfo {author} {\bibfnamefont {Klaus}\
  \bibnamefont {Sengstock}},\ }\bibfield  {title} {\enquote {\bibinfo {title}
  {Oscillations and interactions of dark and dark--bright solitons in
  bose--einstein condensates},}\ }\href {\doibase 10.1038/nphys962} {\bibfield
  {journal} {\bibinfo  {journal} {Nature Physics}\ }\textbf {\bibinfo {volume}
  {4}},\ \bibinfo {pages} {496--501} (\bibinfo {year} {2008})}\BibitemShut
  {NoStop}%
\bibitem [{\citenamefont {Antezza}\ \emph {et~al.}(2007)\citenamefont
  {Antezza}, \citenamefont {Dalfovo}, \citenamefont {Pitaevskii},\ and\
  \citenamefont {Stringari}}]{Antezza2007}%
  \BibitemOpen
  \bibfield  {author} {\bibinfo {author} {\bibfnamefont {Mauro}\ \bibnamefont
  {Antezza}}, \bibinfo {author} {\bibfnamefont {Franco}\ \bibnamefont
  {Dalfovo}}, \bibinfo {author} {\bibfnamefont {Lev~P.}\ \bibnamefont
  {Pitaevskii}}, \ and\ \bibinfo {author} {\bibfnamefont {Sandro}\ \bibnamefont
  {Stringari}},\ }\bibfield  {title} {\enquote {\bibinfo {title} {Dark solitons
  in a superfluid fermi gas},}\ }\href {\doibase 10.1103/PhysRevA.76.043610}
  {\bibfield  {journal} {\bibinfo  {journal} {Phys. Rev. A}\ }\textbf {\bibinfo
  {volume} {76}},\ \bibinfo {pages} {043610} (\bibinfo {year}
  {2007})}\BibitemShut {NoStop}%
\bibitem [{\citenamefont {Scott}\ \emph {et~al.}(2011)\citenamefont {Scott},
  \citenamefont {Dalfovo}, \citenamefont {Pitaevskii},\ and\ \citenamefont
  {Stringari}}]{Robin2011}%
  \BibitemOpen
  \bibfield  {author} {\bibinfo {author} {\bibfnamefont {R.~G.}\ \bibnamefont
  {Scott}}, \bibinfo {author} {\bibfnamefont {F.}~\bibnamefont {Dalfovo}},
  \bibinfo {author} {\bibfnamefont {L.~P.}\ \bibnamefont {Pitaevskii}}, \ and\
  \bibinfo {author} {\bibfnamefont {S.}~\bibnamefont {Stringari}},\ }\bibfield
  {title} {\enquote {\bibinfo {title} {Dynamics of dark solitons in a trapped
  superfluid fermi gas},}\ }\href {\doibase 10.1103/PhysRevLett.106.185301}
  {\bibfield  {journal} {\bibinfo  {journal} {Phys. Rev. Lett.}\ }\textbf
  {\bibinfo {volume} {106}},\ \bibinfo {pages} {185301} (\bibinfo {year}
  {2011})}\BibitemShut {NoStop}%
\bibitem [{\citenamefont {Spuntarelli}\ \emph {et~al.}(2011)\citenamefont
  {Spuntarelli}, \citenamefont {Carr}, \citenamefont {Pieri},\ and\
  \citenamefont {Strinati}}]{Spuntarelli2011}%
  \BibitemOpen
  \bibfield  {author} {\bibinfo {author} {\bibfnamefont {Andrea}\ \bibnamefont
  {Spuntarelli}}, \bibinfo {author} {\bibfnamefont {Lincoln~D}\ \bibnamefont
  {Carr}}, \bibinfo {author} {\bibfnamefont {Pierbiagio}\ \bibnamefont
  {Pieri}}, \ and\ \bibinfo {author} {\bibfnamefont {Giancarlo~C}\ \bibnamefont
  {Strinati}},\ }\bibfield  {title} {\enquote {\bibinfo {title} {Gray solitons
  in a strongly interacting superfluid fermi gas},}\ }\href {\doibase
  10.1088/1367-2630/13/3/035010} {\bibfield  {journal} {\bibinfo  {journal}
  {New Journal of Physics}\ }\textbf {\bibinfo {volume} {13}},\ \bibinfo
  {pages} {035010} (\bibinfo {year} {2011})}\BibitemShut {NoStop}%
\bibitem [{\citenamefont {Liao}\ and\ \citenamefont {Brand}(2011)}]{Liao2011}%
  \BibitemOpen
  \bibfield  {author} {\bibinfo {author} {\bibfnamefont {Renyuan}\ \bibnamefont
  {Liao}}\ and\ \bibinfo {author} {\bibfnamefont {Joachim}\ \bibnamefont
  {Brand}},\ }\bibfield  {title} {\enquote {\bibinfo {title} {Traveling dark
  solitons in superfluid fermi gases},}\ }\href {\doibase
  10.1103/PhysRevA.83.041604} {\bibfield  {journal} {\bibinfo  {journal} {Phys.
  Rev. A}\ }\textbf {\bibinfo {volume} {83}},\ \bibinfo {pages} {041604}
  (\bibinfo {year} {2011})}\BibitemShut {NoStop}%
\bibitem [{\citenamefont {Scott}\ \emph {et~al.}(2012)\citenamefont {Scott},
  \citenamefont {Dalfovo}, \citenamefont {Pitaevskii}, \citenamefont
  {Stringari}, \citenamefont {Fialko}, \citenamefont {Liao},\ and\
  \citenamefont {Brand}}]{Robin2012njp}%
  \BibitemOpen
  \bibfield  {author} {\bibinfo {author} {\bibfnamefont {R~G}\ \bibnamefont
  {Scott}}, \bibinfo {author} {\bibfnamefont {F}~\bibnamefont {Dalfovo}},
  \bibinfo {author} {\bibfnamefont {L~P}\ \bibnamefont {Pitaevskii}}, \bibinfo
  {author} {\bibfnamefont {S}~\bibnamefont {Stringari}}, \bibinfo {author}
  {\bibfnamefont {O}~\bibnamefont {Fialko}}, \bibinfo {author} {\bibfnamefont
  {R}~\bibnamefont {Liao}}, \ and\ \bibinfo {author} {\bibfnamefont
  {J}~\bibnamefont {Brand}},\ }\bibfield  {title} {\enquote {\bibinfo {title}
  {The decay and collisions of dark solitons in superfluid fermi gases},}\
  }\href {\doibase 10.1088/1367-2630/14/2/023044} {\bibfield  {journal}
  {\bibinfo  {journal} {New Journal of Physics}\ }\textbf {\bibinfo {volume}
  {14}},\ \bibinfo {pages} {023044} (\bibinfo {year} {2012})}\BibitemShut
  {NoStop}%
\bibitem [{\citenamefont {Efimkin}\ and\ \citenamefont
  {Galitski}(2015)}]{Efimkin2015}%
  \BibitemOpen
  \bibfield  {author} {\bibinfo {author} {\bibfnamefont {Dmitry~K.}\
  \bibnamefont {Efimkin}}\ and\ \bibinfo {author} {\bibfnamefont {Victor}\
  \bibnamefont {Galitski}},\ }\bibfield  {title} {\enquote {\bibinfo {title}
  {Moving solitons in a one-dimensional fermionic superfluid},}\ }\href
  {\doibase 10.1103/PhysRevA.91.023616} {\bibfield  {journal} {\bibinfo
  {journal} {Phys. Rev. A}\ }\textbf {\bibinfo {volume} {91}},\ \bibinfo
  {pages} {023616} (\bibinfo {year} {2015})}\BibitemShut {NoStop}%
\bibitem [{\citenamefont {Wang}\ \emph {et~al.}(2012)\citenamefont {Wang},
  \citenamefont {Yu}, \citenamefont {Fu}, \citenamefont {Miao}, \citenamefont
  {Huang}, \citenamefont {Chai}, \citenamefont {Zhai},\ and\ \citenamefont
  {Zhang}}]{Wang2012prl}%
  \BibitemOpen
  \bibfield  {author} {\bibinfo {author} {\bibfnamefont {Pengjun}\ \bibnamefont
  {Wang}}, \bibinfo {author} {\bibfnamefont {Zeng-Qiang}\ \bibnamefont {Yu}},
  \bibinfo {author} {\bibfnamefont {Zhengkun}\ \bibnamefont {Fu}}, \bibinfo
  {author} {\bibfnamefont {Jiao}\ \bibnamefont {Miao}}, \bibinfo {author}
  {\bibfnamefont {Lianghui}\ \bibnamefont {Huang}}, \bibinfo {author}
  {\bibfnamefont {Shijie}\ \bibnamefont {Chai}}, \bibinfo {author}
  {\bibfnamefont {Hui}\ \bibnamefont {Zhai}}, \ and\ \bibinfo {author}
  {\bibfnamefont {Jing}\ \bibnamefont {Zhang}},\ }\bibfield  {title} {\enquote
  {\bibinfo {title} {Spin-orbit coupled degenerate fermi gases},}\ }\href
  {\doibase 10.1103/PhysRevLett.109.095301} {\bibfield  {journal} {\bibinfo
  {journal} {Phys. Rev. Lett.}\ }\textbf {\bibinfo {volume} {109}},\ \bibinfo
  {pages} {095301} (\bibinfo {year} {2012})}\BibitemShut {NoStop}%
\bibitem [{\citenamefont {Cheuk}\ \emph {et~al.}(2012)\citenamefont {Cheuk},
  \citenamefont {Sommer}, \citenamefont {Hadzibabic}, \citenamefont {Yefsah},
  \citenamefont {Bakr},\ and\ \citenamefont {Zwierlein}}]{Cheuk2012prl}%
  \BibitemOpen
  \bibfield  {author} {\bibinfo {author} {\bibfnamefont {Lawrence~W.}\
  \bibnamefont {Cheuk}}, \bibinfo {author} {\bibfnamefont {Ariel~T.}\
  \bibnamefont {Sommer}}, \bibinfo {author} {\bibfnamefont {Zoran}\
  \bibnamefont {Hadzibabic}}, \bibinfo {author} {\bibfnamefont {Tarik}\
  \bibnamefont {Yefsah}}, \bibinfo {author} {\bibfnamefont {Waseem~S.}\
  \bibnamefont {Bakr}}, \ and\ \bibinfo {author} {\bibfnamefont {Martin~W.}\
  \bibnamefont {Zwierlein}},\ }\bibfield  {title} {\enquote {\bibinfo {title}
  {Spin-injection spectroscopy of a spin-orbit coupled fermi gas},}\ }\href
  {\doibase 10.1103/PhysRevLett.109.095302} {\bibfield  {journal} {\bibinfo
  {journal} {Phys. Rev. Lett.}\ }\textbf {\bibinfo {volume} {109}},\ \bibinfo
  {pages} {095302} (\bibinfo {year} {2012})}\BibitemShut {NoStop}%
\bibitem [{\citenamefont {Xu}\ \emph {et~al.}(2014)\citenamefont {Xu},
  \citenamefont {Mao}, \citenamefont {Wu},\ and\ \citenamefont
  {Zhang}}]{Xu2014}%
  \BibitemOpen
  \bibfield  {author} {\bibinfo {author} {\bibfnamefont {Yong}\ \bibnamefont
  {Xu}}, \bibinfo {author} {\bibfnamefont {Li}~\bibnamefont {Mao}}, \bibinfo
  {author} {\bibfnamefont {Biao}\ \bibnamefont {Wu}}, \ and\ \bibinfo {author}
  {\bibfnamefont {Chuanwei}\ \bibnamefont {Zhang}},\ }\bibfield  {title}
  {\enquote {\bibinfo {title} {Dark solitons with majorana fermions in
  spin-orbit-coupled fermi gases},}\ }\href {\doibase
  10.1103/PhysRevLett.113.130404} {\bibfield  {journal} {\bibinfo  {journal}
  {Phys. Rev. Lett.}\ }\textbf {\bibinfo {volume} {113}},\ \bibinfo {pages}
  {130404} (\bibinfo {year} {2014})}\BibitemShut {NoStop}%
\bibitem [{\citenamefont {Liu}(2015)}]{Liu2015}%
  \BibitemOpen
  \bibfield  {author} {\bibinfo {author} {\bibfnamefont {Xia-Ji}\ \bibnamefont
  {Liu}},\ }\bibfield  {title} {\enquote {\bibinfo {title} {Soliton-induced
  majorana fermions in a one-dimensional atomic topological superfluid},}\
  }\href {\doibase 10.1103/PhysRevA.91.023610} {\bibfield  {journal} {\bibinfo
  {journal} {Phys. Rev. A}\ }\textbf {\bibinfo {volume} {91}},\ \bibinfo
  {pages} {023610} (\bibinfo {year} {2015})}\BibitemShut {NoStop}%
\bibitem [{\citenamefont {Zou}\ \emph {et~al.}(2016)\citenamefont {Zou},
  \citenamefont {Brand}, \citenamefont {Liu},\ and\ \citenamefont
  {Hu}}]{Zou2016}%
  \BibitemOpen
  \bibfield  {author} {\bibinfo {author} {\bibfnamefont {Peng}\ \bibnamefont
  {Zou}}, \bibinfo {author} {\bibfnamefont {Joachim}\ \bibnamefont {Brand}},
  \bibinfo {author} {\bibfnamefont {Xia-Ji}\ \bibnamefont {Liu}}, \ and\
  \bibinfo {author} {\bibfnamefont {Hui}\ \bibnamefont {Hu}},\ }\bibfield
  {title} {\enquote {\bibinfo {title} {Traveling majorana solitons in a
  low-dimensional spin-orbit-coupled fermi superfluid},}\ }\href {\doibase
  10.1103/PhysRevLett.117.225302} {\bibfield  {journal} {\bibinfo  {journal}
  {Phys. Rev. Lett.}\ }\textbf {\bibinfo {volume} {117}},\ \bibinfo {pages}
  {225302} (\bibinfo {year} {2016})}\BibitemShut {NoStop}%
\bibitem [{\citenamefont {Bongs}\ \emph {et~al.}(2003)\citenamefont {Bongs},
  \citenamefont {Burger}, \citenamefont {Hellweg}, \citenamefont {Kottke},
  \citenamefont {Dettmer}, \citenamefont {Rinkleff}, \citenamefont
  {Cacciapuoti}, \citenamefont {Arlt}, \citenamefont {Sengstock},\ and\
  \citenamefont {Ertmer}}]{Bongs2003}%
  \BibitemOpen
  \bibfield  {author} {\bibinfo {author} {\bibfnamefont {K}~\bibnamefont
  {Bongs}}, \bibinfo {author} {\bibfnamefont {S}~\bibnamefont {Burger}},
  \bibinfo {author} {\bibfnamefont {D}~\bibnamefont {Hellweg}}, \bibinfo
  {author} {\bibfnamefont {M}~\bibnamefont {Kottke}}, \bibinfo {author}
  {\bibfnamefont {S}~\bibnamefont {Dettmer}}, \bibinfo {author} {\bibfnamefont
  {T}~\bibnamefont {Rinkleff}}, \bibinfo {author} {\bibfnamefont
  {L}~\bibnamefont {Cacciapuoti}}, \bibinfo {author} {\bibfnamefont
  {J}~\bibnamefont {Arlt}}, \bibinfo {author} {\bibfnamefont {K}~\bibnamefont
  {Sengstock}}, \ and\ \bibinfo {author} {\bibfnamefont {W}~\bibnamefont
  {Ertmer}},\ }\bibfield  {title} {\enquote {\bibinfo {title} {Spectroscopy of
  dark soliton states in bose~einstein condensates},}\ }\href {\doibase
  10.1088/1464-4266/5/2/369} {\bibfield  {journal} {\bibinfo  {journal}
  {Journal of Optics B: Quantum and Semiclassical Optics}\ }\textbf {\bibinfo
  {volume} {5}},\ \bibinfo {pages} {S124--S130} (\bibinfo {year}
  {2003})}\BibitemShut {NoStop}%
\bibitem [{\citenamefont {Burger}\ \emph {et~al.}(1999)\citenamefont {Burger},
  \citenamefont {Bongs}, \citenamefont {Dettmer}, \citenamefont {Ertmer},
  \citenamefont {Sengstock}, \citenamefont {Sanpera}, \citenamefont
  {Shlyapnikov},\ and\ \citenamefont {Lewenstein}}]{Burger1999}%
  \BibitemOpen
  \bibfield  {author} {\bibinfo {author} {\bibfnamefont {S.}~\bibnamefont
  {Burger}}, \bibinfo {author} {\bibfnamefont {K.}~\bibnamefont {Bongs}},
  \bibinfo {author} {\bibfnamefont {S.}~\bibnamefont {Dettmer}}, \bibinfo
  {author} {\bibfnamefont {W.}~\bibnamefont {Ertmer}}, \bibinfo {author}
  {\bibfnamefont {K.}~\bibnamefont {Sengstock}}, \bibinfo {author}
  {\bibfnamefont {A.}~\bibnamefont {Sanpera}}, \bibinfo {author} {\bibfnamefont
  {G.~V.}\ \bibnamefont {Shlyapnikov}}, \ and\ \bibinfo {author} {\bibfnamefont
  {M.}~\bibnamefont {Lewenstein}},\ }\bibfield  {title} {\enquote {\bibinfo
  {title} {Dark solitons in bose-einstein condensates},}\ }\href {\doibase
  10.1103/PhysRevLett.83.5198} {\bibfield  {journal} {\bibinfo  {journal}
  {Phys. Rev. Lett.}\ }\textbf {\bibinfo {volume} {83}},\ \bibinfo {pages}
  {5198--5201} (\bibinfo {year} {1999})}\BibitemShut {NoStop}%
\bibitem [{\citenamefont {Yefsah}\ \emph {et~al.}(2013)\citenamefont {Yefsah},
  \citenamefont {Sommer}, \citenamefont {Ku}, \citenamefont {Cheuk},
  \citenamefont {Ji}, \citenamefont {Bakr},\ and\ \citenamefont
  {Zwierlein}}]{Yefsah2013}%
  \BibitemOpen
  \bibfield  {author} {\bibinfo {author} {\bibfnamefont {Tarik}\ \bibnamefont
  {Yefsah}}, \bibinfo {author} {\bibfnamefont {Ariel~T.}\ \bibnamefont
  {Sommer}}, \bibinfo {author} {\bibfnamefont {Mark J.~H.}\ \bibnamefont {Ku}},
  \bibinfo {author} {\bibfnamefont {Lawrence~W.}\ \bibnamefont {Cheuk}},
  \bibinfo {author} {\bibfnamefont {Wenjie}\ \bibnamefont {Ji}}, \bibinfo
  {author} {\bibfnamefont {Waseem~S.}\ \bibnamefont {Bakr}}, \ and\ \bibinfo
  {author} {\bibfnamefont {Martin~W.}\ \bibnamefont {Zwierlein}},\ }\bibfield
  {title} {\enquote {\bibinfo {title} {Heavy solitons in a fermionic
  superfluid},}\ }\href {\doibase 10.1038/nature12338} {\bibfield  {journal}
  {\bibinfo  {journal} {Nature}\ }\textbf {\bibinfo {volume} {499}},\ \bibinfo
  {pages} {426--430} (\bibinfo {year} {2013})}\BibitemShut {NoStop}%
\bibitem [{\citenamefont {Bulgac}\ \emph {et~al.}(2014)\citenamefont {Bulgac},
  \citenamefont {Forbes}, \citenamefont {Kelley}, \citenamefont {Roche},\ and\
  \citenamefont {Wlaz\l{}owski}}]{Bulgac2014}%
  \BibitemOpen
  \bibfield  {author} {\bibinfo {author} {\bibfnamefont {Aurel}\ \bibnamefont
  {Bulgac}}, \bibinfo {author} {\bibfnamefont {Michael~McNeil}\ \bibnamefont
  {Forbes}}, \bibinfo {author} {\bibfnamefont {Michelle~M.}\ \bibnamefont
  {Kelley}}, \bibinfo {author} {\bibfnamefont {Kenneth~J.}\ \bibnamefont
  {Roche}}, \ and\ \bibinfo {author} {\bibfnamefont {Gabriel}\ \bibnamefont
  {Wlaz\l{}owski}},\ }\bibfield  {title} {\enquote {\bibinfo {title} {Quantized
  superfluid vortex rings in the unitary fermi gas},}\ }\href {\doibase
  10.1103/PhysRevLett.112.025301} {\bibfield  {journal} {\bibinfo  {journal}
  {Phys. Rev. Lett.}\ }\textbf {\bibinfo {volume} {112}},\ \bibinfo {pages}
  {025301} (\bibinfo {year} {2014})}\BibitemShut {NoStop}%
\bibitem [{\citenamefont {Donadello}\ \emph {et~al.}(2014)\citenamefont
  {Donadello}, \citenamefont {Serafini}, \citenamefont {Tylutki}, \citenamefont
  {Pitaevskii}, \citenamefont {Dalfovo}, \citenamefont {Lamporesi},\ and\
  \citenamefont {Ferrari}}]{Donadello2014}%
  \BibitemOpen
  \bibfield  {author} {\bibinfo {author} {\bibfnamefont {Simone}\ \bibnamefont
  {Donadello}}, \bibinfo {author} {\bibfnamefont {Simone}\ \bibnamefont
  {Serafini}}, \bibinfo {author} {\bibfnamefont {Marek}\ \bibnamefont
  {Tylutki}}, \bibinfo {author} {\bibfnamefont {Lev~P.}\ \bibnamefont
  {Pitaevskii}}, \bibinfo {author} {\bibfnamefont {Franco}\ \bibnamefont
  {Dalfovo}}, \bibinfo {author} {\bibfnamefont {Giacomo}\ \bibnamefont
  {Lamporesi}}, \ and\ \bibinfo {author} {\bibfnamefont {Gabriele}\
  \bibnamefont {Ferrari}},\ }\bibfield  {title} {\enquote {\bibinfo {title}
  {Observation of solitonic vortices in bose-einstein condensates},}\ }\href
  {\doibase 10.1103/PhysRevLett.113.065302} {\bibfield  {journal} {\bibinfo
  {journal} {Phys. Rev. Lett.}\ }\textbf {\bibinfo {volume} {113}},\ \bibinfo
  {pages} {065302} (\bibinfo {year} {2014})}\BibitemShut {NoStop}%
\bibitem [{\citenamefont {Sacha}\ and\ \citenamefont
  {Delande}(2014)}]{Sacha2014}%
  \BibitemOpen
  \bibfield  {author} {\bibinfo {author} {\bibfnamefont {Krzysztof}\
  \bibnamefont {Sacha}}\ and\ \bibinfo {author} {\bibfnamefont {Dominique}\
  \bibnamefont {Delande}},\ }\bibfield  {title} {\enquote {\bibinfo {title}
  {Proper phase imprinting method for a dark soliton excitation in a superfluid
  fermi mixture},}\ }\href {\doibase 10.1103/PhysRevA.90.021604} {\bibfield
  {journal} {\bibinfo  {journal} {Phys. Rev. A}\ }\textbf {\bibinfo {volume}
  {90}},\ \bibinfo {pages} {021604} (\bibinfo {year} {2014})}\BibitemShut
  {NoStop}%
\bibitem [{\citenamefont {Paw{\l}owski}\ and\ \citenamefont
  {Rz{\k{a}}{\.{z}}ewski}(2015)}]{Pawlowski2015njp}%
  \BibitemOpen
  \bibfield  {author} {\bibinfo {author} {\bibfnamefont {Krzysztof}\
  \bibnamefont {Paw{\l}owski}}\ and\ \bibinfo {author} {\bibfnamefont
  {Kazimierz}\ \bibnamefont {Rz{\k{a}}{\.{z}}ewski}},\ }\bibfield  {title}
  {\enquote {\bibinfo {title} {Dipolar dark solitons},}\ }\href {\doibase
  10.1088/1367-2630/17/10/105006} {\bibfield  {journal} {\bibinfo  {journal}
  {New Journal of Physics}\ }\textbf {\bibinfo {volume} {17}},\ \bibinfo
  {pages} {105006} (\bibinfo {year} {2015})}\BibitemShut {NoStop}%
\bibitem [{\citenamefont {Liao}\ \emph {et~al.}(2010)\citenamefont {Liao},
  \citenamefont {Rittner}, \citenamefont {Paprotta}, \citenamefont {Li},
  \citenamefont {Partridge}, \citenamefont {Hulet}, \citenamefont {Baur},\ and\
  \citenamefont {Mueller}}]{Liao2010}%
  \BibitemOpen
  \bibfield  {author} {\bibinfo {author} {\bibfnamefont {Yean-an}\ \bibnamefont
  {Liao}}, \bibinfo {author} {\bibfnamefont {Ann Sophie~C.}\ \bibnamefont
  {Rittner}}, \bibinfo {author} {\bibfnamefont {Tobias}\ \bibnamefont
  {Paprotta}}, \bibinfo {author} {\bibfnamefont {Wenhui}\ \bibnamefont {Li}},
  \bibinfo {author} {\bibfnamefont {Guthrie~B.}\ \bibnamefont {Partridge}},
  \bibinfo {author} {\bibfnamefont {Randall~G.}\ \bibnamefont {Hulet}},
  \bibinfo {author} {\bibfnamefont {Stefan~K.}\ \bibnamefont {Baur}}, \ and\
  \bibinfo {author} {\bibfnamefont {Erich~J.}\ \bibnamefont {Mueller}},\
  }\bibfield  {title} {\enquote {\bibinfo {title} {Spin-imbalance in a
  one-dimensional fermi gas},}\ }\href {\doibase 10.1038/nature09393}
  {\bibfield  {journal} {\bibinfo  {journal} {Nature}\ }\textbf {\bibinfo
  {volume} {467}},\ \bibinfo {pages} {567--569} (\bibinfo {year}
  {2010})}\BibitemShut {NoStop}%
\bibitem [{\citenamefont {Liu}\ \emph {et~al.}(2007{\natexlab{a}})\citenamefont
  {Liu}, \citenamefont {Hu},\ and\ \citenamefont {Drummond}}]{Liu2007pra}%
  \BibitemOpen
  \bibfield  {author} {\bibinfo {author} {\bibfnamefont {Xia-Ji}\ \bibnamefont
  {Liu}}, \bibinfo {author} {\bibfnamefont {Hui}\ \bibnamefont {Hu}}, \ and\
  \bibinfo {author} {\bibfnamefont {Peter~D.}\ \bibnamefont {Drummond}},\
  }\bibfield  {title} {\enquote {\bibinfo {title}
  {Fulde-ferrell-larkin-ovchinnikov states in one-dimensional spin-polarized
  ultracold atomic fermi gases},}\ }\href {\doibase 10.1103/PhysRevA.76.043605}
  {\bibfield  {journal} {\bibinfo  {journal} {Phys. Rev. A}\ }\textbf {\bibinfo
  {volume} {76}},\ \bibinfo {pages} {043605} (\bibinfo {year}
  {2007}{\natexlab{a}})}\BibitemShut {NoStop}%
\bibitem [{\citenamefont {Liu}\ \emph {et~al.}(2008)\citenamefont {Liu},
  \citenamefont {Hu},\ and\ \citenamefont {Drummond}}]{Liu2008pra}%
  \BibitemOpen
  \bibfield  {author} {\bibinfo {author} {\bibfnamefont {Xia-Ji}\ \bibnamefont
  {Liu}}, \bibinfo {author} {\bibfnamefont {Hui}\ \bibnamefont {Hu}}, \ and\
  \bibinfo {author} {\bibfnamefont {Peter~D.}\ \bibnamefont {Drummond}},\
  }\bibfield  {title} {\enquote {\bibinfo {title} {Finite-temperature phase
  diagram of a spin-polarized ultracold fermi gas in a highly elongated
  harmonic trap},}\ }\href {\doibase 10.1103/PhysRevA.78.023601} {\bibfield
  {journal} {\bibinfo  {journal} {Phys. Rev. A}\ }\textbf {\bibinfo {volume}
  {78}},\ \bibinfo {pages} {023601} (\bibinfo {year} {2008})}\BibitemShut
  {NoStop}%
\bibitem [{\citenamefont {Liu}\ \emph {et~al.}(2007{\natexlab{b}})\citenamefont
  {Liu}, \citenamefont {Hu},\ and\ \citenamefont {Drummond}}]{Liu2007}%
  \BibitemOpen
  \bibfield  {author} {\bibinfo {author} {\bibfnamefont {Xia-Ji}\ \bibnamefont
  {Liu}}, \bibinfo {author} {\bibfnamefont {Hui}\ \bibnamefont {Hu}}, \ and\
  \bibinfo {author} {\bibfnamefont {Peter~D.}\ \bibnamefont {Drummond}},\
  }\bibfield  {title} {\enquote {\bibinfo {title} {Mean-field thermodynamics of
  a spin-polarized spherically trapped fermi gas at unitarity},}\ }\href
  {\doibase 10.1103/PhysRevA.75.023614} {\bibfield  {journal} {\bibinfo
  {journal} {Phys. Rev. A}\ }\textbf {\bibinfo {volume} {75}},\ \bibinfo
  {pages} {023614} (\bibinfo {year} {2007}{\natexlab{b}})}\BibitemShut
  {NoStop}%
\bibitem [{\citenamefont {Trubko}\ \emph {et~al.}(2017)\citenamefont {Trubko},
  \citenamefont {Gregoire}, \citenamefont {Holmgren},\ and\ \citenamefont
  {Cronin}}]{Raisa2017}%
  \BibitemOpen
  \bibfield  {author} {\bibinfo {author} {\bibfnamefont {Raisa}\ \bibnamefont
  {Trubko}}, \bibinfo {author} {\bibfnamefont {Maxwell~D.}\ \bibnamefont
  {Gregoire}}, \bibinfo {author} {\bibfnamefont {William~F.}\ \bibnamefont
  {Holmgren}}, \ and\ \bibinfo {author} {\bibfnamefont {Alexander~D.}\
  \bibnamefont {Cronin}},\ }\bibfield  {title} {\enquote {\bibinfo {title}
  {Potassium tune-out-wavelength measurement using atom interferometry and a
  multipass optical cavity},}\ }\href {\doibase 10.1103/PhysRevA.95.052507}
  {\bibfield  {journal} {\bibinfo  {journal} {Phys. Rev. A}\ }\textbf {\bibinfo
  {volume} {95}},\ \bibinfo {pages} {052507} (\bibinfo {year}
  {2017})}\BibitemShut {NoStop}%
\bibitem [{\citenamefont {Jiang}\ \emph {et~al.}(2020)\citenamefont {Jiang},
  \citenamefont {Li}, \citenamefont {Wang}, \citenamefont {Dong},\ and\
  \citenamefont {Wu}}]{Jun2020}%
  \BibitemOpen
  \bibfield  {author} {\bibinfo {author} {\bibfnamefont {Jun}\ \bibnamefont
  {Jiang}}, \bibinfo {author} {\bibfnamefont {Xian-Jun}\ \bibnamefont {Li}},
  \bibinfo {author} {\bibfnamefont {Xia}\ \bibnamefont {Wang}}, \bibinfo
  {author} {\bibfnamefont {Chen-Zhong}\ \bibnamefont {Dong}}, \ and\ \bibinfo
  {author} {\bibfnamefont {Z.~W.}\ \bibnamefont {Wu}},\ }\bibfield  {title}
  {\enquote {\bibinfo {title} {Tune-out wavelengths of the hyperfine components
  of the ground level of $^{133}\mathrm{Cs}$ atoms},}\ }\href {\doibase
  10.1103/PhysRevA.102.042823} {\bibfield  {journal} {\bibinfo  {journal}
  {Phys. Rev. A}\ }\textbf {\bibinfo {volume} {102}},\ \bibinfo {pages}
  {042823} (\bibinfo {year} {2020})}\BibitemShut {NoStop}%
\end{thebibliography}%

\end{document}